# The Unusual Suppression of Superconducting Transition Temperature in Double-Doping 2H-NbSe$_2$

*Dong Yan[1], Yishi Lin[2], Guohua Wang[3], Zhen Zhu[3], Shu Wang[1], Lei Shi[1], Yuan He[1], Man-Rong Li[4], Hao Zheng[3], Jie Ma[3], Jinfeng Jia[3], Yihua Wang[2], Huixia Luo[1,5]\**

[1]School of Material Science and Engineering, Sun Yat-Sen University, No. 135, Xingang Xi Road, Guangzhou, 510275, P. R. China
[2]Department of Physics, Fudan University, Shanghai, 200433, China
[3]Department of Physics and Astronomy, Shanghai Jiao Tong University, Shanghai 200240, China
[4]School of Chemistry, Sun Yat-Sen University, No. 135, Xingang Xi Road, Guangzhou, 510275, China
[5]Key Lab Polymer Composite & Functional Materials, Sun Yat-Sen University, No. 135, Xingang Xi Road, Guangzhou, 510275, P. R. China

E-mail: luohx7@mail.sysu.edu.cn

**ABSTRACT**

2H-NbSe$_2$ is one of the most widely researched transition metal dichalcogenide (TMD) superconductors, which undergoes charge-density wave (CDW) transition at $T_{CDW}$ about 33 K and superconducting transition at $T_c$ of 7.3 K. To explore the relation between its superconductivity and Fermi surface nesting, we combined S substitution with Cu intercalation in 2H-NbSe$_2$ to make Cu$_x$NbSe$_{2-y}$S$_y$ ($0 \leq x = y \leq 0.1$). Upon systematic substitution of S and intercalation of Cu ions into 2H-NbSe$_2$, we found that when the Cu and S contents ($x = y \geq 0.06$) increases, the $T_c$ decreases in Cu$_x$NbSe$_{2-y}$S$_y$. While at higher $x$ and $y$ values, $T_c$ keeps a constant value near 2 K, which is not commonly observed for a layered TMD. For comparison, we found the simultaneous substitution of Nb by Cu and Se by S in Cu$_x$Nb$_{1-x}$Se$_{2-y}$S$_y$ ($0 \leq x = y \leq 0.06$) lowered the $T_c$ substantially faster. We construct a superconducting phase diagrams for our double-doping compounds in contrast with the related single-ions doping systems.

## INTRODUCTION

Transition metal dichalcogenides (TMDs) with 1T (octahedrally coordinated) and 2H (trigonal prismatically coordinated) structures have attracted abundant research for decades due their various novel properties, *i. e* superconductivity and charge density wave (CDW) and topological semimetal states. [1-8] 2H-NbSe$_2$ has been researched as the most famous layered superconducting TMD materials bearing CDW state with the superconducting transition temperature ($T_c$) of 7.3 K and a quasi-two-dimensional incommensurate CDW transition temperature (T$_{ICDW}$) of ~ 33 K. [9] The co-existence of superconductivity and CDW state and its superconducting mechanism have made it a research hotspot. [10-13] Even if 2H-NbSe$_2$ has been taken for representative superconductor for many years, it has also been recently found that 2H-NbSe$_2$ is a multiband superconductor with some similarities to that of MgB$_2$ superconductor. [14-16] The relationship between its superconductivity and Fermi surface nesting is still under fierce debate.

In order to figure out this debate, numerous theoretical and experimental articles have been concentrated researching on its behavior. In the experimental case, researchers usually used the familiar and effective ways to tune the superconducting and CDW transition temperatures of this class, including chemical doping (*e.g.*, refs 17-22), adopting high pressure (*e.g.*, refs 23-26) and gating.[27] Especially, a large number of research papers focusing on tuning its superconducting and CDW behavior by chemical doping have been published. So far, all the inorganic doping 2H-NbSe$_2$ system shows lower $T_c$, comparing with the pristine 2H-NbSe$_2$. For example, the 3$d$ transition metals such as Fe, Ni, Co, Zn, or Al have been used to intercalate into 2H-NbSe$_2$ and resulted in remarkable $T_c$ drop. [28-29] More recently, we studied the effect of Cu intercalation of 2H-NbSe$_2$ and found an unusual *S*-shaped superconducting phase diagram in 2H-Cu$_x$NbSe$_2$.[20] In most of the superconductors, the transition temperature $T_c$ decreases with increasing disorder.[30] However, in the 2H-Cu$_x$NbSe$_2$ system, as is generally observed the $T_c$ will drop off with an increased doping value when introduce "impurities" into an optimal superconductor, whereas the mode that $T_c$ declined under such instance is novel: an *S*-shaped superconducting phase diagram was obtained, near $x = 0.03$ with an inflection point and appearing a leveling off of the $T_c$ about 3 K (usual value of layered TMD) at high doing content. In contrast, there is minor change in $T_c$ of the S substitution for Se in 2H-NbSe$_2$ system. [31-32]

Superconducting state coexists with a CDW state or occurs in proximity of such a state in most metallic TMDs. It has been thought for decades that the occurrence of a CDW state was associated with superconductivity. 2H-NbS$_2$ is isostructural and isoelectronic to 2H-NbSe$_2$ and with a $T_c$ of 6.05 K, but no CDW order occurs, [33] which does not accord with the diversity of electronic instabilities in the TMDs. Pristine 2H-NbSe$_2$ and 2H-NbS$_2$ are both multiband superconductors, and the former one has a CDW state but the latter one doesn't. [33-35] On the other hand, S doped 2H-NbSe$_2$ slightly affects the $T_c$ but unusual $S$-shaped superconducting phase diagram is observed when Cu ions intercalated into 2H-NbSe$_2$. [20, 31-32] 2H-NbSe$_2$ and 2H-NbS$_2$ thus open a door to the study of the interaction of CDW and multiband superconductivity phenomena in the bulk TMDs. Tuning the physical properties of CDW and superconducting states by chemical doping including substitution and intercalation is a common tool in the condensed matter field. In order to further study the interplay of the multiband superconductivity and CDW state in the TMDs, we proposed to combine with S substitution and Cu substitution/intercalation of 2H-NbSe$_2$. To the best of our knowledge, there was no work about double doping in 2H-NbSe$_2$ to date.

In this article we report the double doping Cu and S into 2H-NbSe$_2$ to form Cu$_x$Nb$_{1-x}$Se$_{2-y}$S$_y$ ($x = y = 0.06$) and Cu$_x$NbSe$_{2-y}$S$_y$ ($0 \leq x = y \leq 0.1$), respectively, with remaining the 2H structure (see **Fig. 1A, B**). The material is multiple phases with $x > 0.1$ in Cu$_x$NbSe$_{2-y}$S$_y$ and $x > 0.06$ in Cu$_x$Nb$_{1-x}$Se$_{2-y}$S$_y$, and thus corresponding $T_c$ cannot be reliably measured. Magnetic susceptibilities, resistivities, and heat capacities measurements were performed to characterize Cu$_x$NbSe$_{2-y}$S$_y$ polycrystalline samples systematically. For comparison, magnetic susceptibilities of Cu$_x$Nb$_{1-x}$Se$_{2-y}$S$_y$ polycrystalline samples were also measured. The results signify that the $T_c$ value has a diminution with increasing Cu and S content in both Cu$_x$Nb$_{1-x}$Se$_{2-y}$S$_y$ and

$Cu_xNbSe_{2-y}S_y$ case, which was usually observed when introduce "foreign matter" into the pristine 2H-$NbSe_2$ superconductors. However, we obtained an *S*-shaped superconducting phase diagram with $T_c$ *vs.* $x$ in the $Cu_xNbSe_{2-y}S_y$ case, having an inflexion at $x = 0.06$ and tending to smooth at high $x$ value, from original $T_c$ about 7.3 K in 2H-$NbSe_2$ to low $T_c$ value near 2 K, which is similar to the single doping case $Cu_xNbSe_2$. Besides, a half arc-shaped superconducting phase diagram with $T_c$ *vs.* $x$ was observed in the $Cu_xNb_{1-x}Se_{2-y}S_y$ case. The superconducting state competes with CDW state in the layered $Cu_xNbSe_{2-y}S_y$ TMDs at low temperatures. The low-temperature scanning tunneling microscopy (STM) study reveals the formation of the 2 × 2 commensurate CDWs on the sample's surface and with tiny variety in *q* vector of the charge density wave state arisen from the Cu intercalation and S substitution, weaken but not exterminate of the correlation of the CDW. The difference between other doped 2H-$NbSe_2$ materials such as $NbSe_{2-x}S_x$, $Fe_xNbSe_2$, $Cu_xNbSe_2$, and $Cu_xNb_{1-x}Se_{2-y}S_y$ in the superconducting phase diagram revealed its uncommon character.

**EXPERIMENTAL SECTION**

$Cu_xNbSe_{2-y}S_y$ ($0 \leq x = y \leq 0.1$) and $Cu_xNb_{1-x}Se_{2-y}S_y$ ($x = y = 0.06$) polycrystalline samples were synthesized *via* solid state method. Firstly, the mixture of regent of Cu (99.9%), S (99.999%), Se (99.999%) and Nb (99.9%) in corresponding stoichiometric ratios were heated to 850 °C at a rate of 2 °C/min and held for 5 days in sealed vacuum silica glass tubes. Then the as-prepared polycrystalline samples were reground, re-pelletized, and sintered again *via* a rate of 5 °C/min to 850 °C and held for 2 days. The selected compositions single crystals were grown by the chemical vapor transport (CVT) method, using iodine as a transport agent. The $Cu_xNbSe_{2-y}S_y$ ($0 \leq x = y \leq 0.1$) as-prepared powders were mixed with iodine at quality ratio 20:1, then heated sealed vacuum silica tubes for 7 days in a two-zone furnace, where the growth and source zones temperatures were 625 and 725 °C, respectively. We also adopt the same method to synthesize the powder and single crystal sample of $Cu_xNb_{1-x}Se_{2-y}S_y$ ($0 \leq x = y \leq 0.06$).

The morphologies of the single crystals were studied by scanning electron microscopy (SEM, Quanta 400F, Oxford) operated at 20 kV. The element distributions and element compositions of single crystals of $Cu_xNbSe_{2-y}S_y$ were investigated by energy dispersive X-ray spectroscopy (EDXS). The $Cu_{0.06}NbSe_{1.71}S_{0.08}$ sample was also characterized by a low-temperature STM (USM-1600, Unisoku) with an ultrahigh vacuum (base pressure ~1 × 10-10 torr) at $T$ = 4.8 K. The sample was cleaved in situ at room temperature and then transferred into the STM head immediately. Electrochemically etched tungsten tips were used after heating and silver decoration.

We adopt powder X-ray diffraction (PXRD) with Bruker D8 Advance ECO equipping LYNXEYE-XE detector and Cu Kα radiation to determine the phase purity of polycrystalline samples. The unit cell parameters were determined by profile fitting the powder diffraction data with the FULLPROF diffraction suite with Thompson-Cox-Hastings pseudo-Voigt peak shapes.[36] The temperature dependent electrical resistivity (4-point means), magnetic susceptibility and heat capacity of polycrystalline samples were measured by EverCool II Quantum Design Physical Property Measurement System (PPMS). The materials show no air-sensitivity sign in the period of measurements. $T_c$s were taken from the intersection of the extrapolations of the normal state and the steepest slope of the superconducting transition region susceptibility for magnetic susceptibility data; the midpoint of the resistivity transitions was referred from resistivities; meanwhile the $T_c$s obtained from the equal area construction method for heat capacity data.

**RESULTS AND DISCUSSION**

**Figs. 1A-F** show the unit cell parameters and PXRD pattern for $Cu_xNbSe_{2-y}S_y$ (0 ≤ $x$ = $y$ ≤ 0.1) and $Cu_xNb_{1-x}Se_{2-y}S_y$ (0 ≤ $x$ = $y$ ≤ 0.06). The results show that the powder samples obtained was pure. The doping limit for substitution of S and intercalated Cu into 2H-NbSe$_2$ is $x$ = 0.1. However, the limit for Cu and S co-substitution into 2H-NbSe$_2$ is x = 0.06. Impurities will appear as shown in Fig. 1A when the doping content over the limit. The unit cell parameters $c$ for both $Cu_xNbSe_{2-y}S_y$ (0 ≤ $x$ = $y$ ≤

0.1) and $Cu_xNb_{1-x}Se_{2-y}S_y$ ($0 \leq x = y \leq 0.06$) increases linearly with increasing Cu and S content within the solid solution. $c$ augments linearly from 12.5680(6) ($x = 0.00$) to 12.6166(3) ($x = 0.10$) in $Cu_xNbSe_{2-y}S_y$ and 12.5680(6) ($x = 0.00$) to 12.5853(3) ($x = 0.06$) in $Cu_xNb_{1-x}Se_{2-y}S_y$, respectively, with Vegard's law type behavior. Inset of **Figs. 1A, B** show the position of peak (002) movement to left with more doping content. This also demonstrates that $c$ will increases with augment of doping (**Fig. 1F**). **Figs. 1C, D** show the refinement results of powder 2H-$Cu_{0.04}Nb_{0.96}Se_{1.96}S_{0.04}$ and $Cu_{0.04}NbSe_{1.96}S_{0.04}$. These reflections can be indexed in $P6_3/mmc$ space group and the lattice parameters derived to be $a = 3.4480(3)$ Å and $c = 12.5785(9)$ Å for $Cu_{0.04}Nb_{0.96}Se_{1.96}S_{0.04}$, $a = 3.4457(3)$ Å and $c = 12.5859(9)$ Å for $Cu_{0.04}NbSe_{1.96}S_{0.04}$. The $Cu_xNbSe_{2-y}S_y$ crystal structure shows the central Nb in trigonal prismatically coordinated by randomly occupied chalcogen (S and Se) to form $MX_2$ layer with Cu intercalating between these layers (**Fig. 1D**). The $Cu_xNb_{1-x}Se_{2-y}S_y$ crystal structure shows the central Nb with Cu-doping in trigonal prismatically coordinated by randomly occupied chalcogen (S and Se) to form $MX_2$ layer (**Fig. 1C**). **Figs. S1A-C** show the SEM and EDXS images of $Cu_{0.06}NbSe_{1.71}S_{0.08}$ single crystal. From **Fig. S1**, we can confirm that the composition of the single crystal is $Cu_{0.06}NbSe_{1.71}S_{0.08}$, which has a layer structure and will be used for scanning tunneling microscopy (STM) measurements later.

**Figs. 2A, B** show the temperature dependence of magnetic susceptibility under applied magnetic field of 20 Oe for $Cu_xNb_{1-x}Se_{2-y}S_y$ ($0 \leq x = y \leq 0.06$) and $Cu_xNbSe_{2-y}S_y$ ($0 \leq x = y \leq 0.1$) in the temperature range 2 - 8 K, respectively. The measurements are performed under zero field cooling. **Fig. 2A** shows the rapid downtrend for $Cu_xNb_{1-x}Se_{2-y}S_y$ ($0 \leq x = y \leq 0.06$), the $T_c$ almost drops to 3.5 K when $x = 0.02$. As shown in **Fig. 2B**, in the case of $Cu_xNbSe_{2-y}S_y$ ($0 \leq x = y \leq 0.1$), $T_c$ also decreases with the increase of the content of intercalation of Cu and substitution of S. We next adopt systematic research on $Cu_xNbSe_{2-y}S_y$. We further got the test on $Cu_xNbSe_{2-y}S_y$ polycrystalline samples after pelletized and sintered for temperature dependence of the normalized electrical resistivities ($\rho/\rho_{300K}$), as shown in **Fig. 2C**. We can easily find the obvious, sharp drop of $\rho(T)$ curves which represent the onset of superconductivity at low temperatures (**Fig. 2D**). The $T_c$s declined with the increasing doping content. Also, this tendency is also apparently found by the susceptibility data of $Cu_xNbSe_{2-y}S_y$ (**Fig. 2B**) - the superconducting state moves to lower temperatures systematically with the increasing doping content in $Cu_xNbSe_{2-y}S_y$ which indicated by the negative magnetic susceptibility. The $Cu_xNbSe_{2-y}S_y$ samples show a metallic temperature dependence between 8 K to 300 K ($d\rho/dT > 0$). Also, at low temperature the $d\rho/dT$ vs. $T$ curve shows peaks at the corresponding $T_c$s (inset of **Fig. 2C**).

In order to get more data of the superconductivity and electronic properties of $Cu_xNbSe_{2-y}S_y$ sosoloid, we adopt heat capacity measurements on their polycrystalline samples. **Fig. 3** shows the temperature dependent zero-field heat capacity, that is $C_p/T$ vs. $T$, for $Cu_{0.02}NbSe_{1.98}S_{0.02}$ sample. The heat capacity curve displays a sharp specific heat jump which indicates the superconducting transition temperature for this material. The obtained $T_c$ from heat capacity is highly consistent with the $T_c$ obtained by the $\chi(T)$ and $\rho(T)$ tests. The normal state of specific heat at high temperatures of zero magnetic field obey the relation of $C_p/T = \beta T^2 + \gamma$, where $\beta$ and $\gamma$ describe the phonon and electronic contributions to the heat capacity, respectively. We got the values of $\beta$ and $\gamma$ (inset of **Figs. 3**) from fitting the data obtained at 0 T field. The normalized specific heat jump value $\Delta C/\gamma T_c$ obtained from the data (**Figs. 3**) was 2.16 for $Cu_{0.02}NbSe_{1.98}S_{0.02}$ which was much higher than the Bardeen-Cooper-Schrieffer (BCS) weak-coupling limit value (1.43). Then we estimate the Debye temperature by the formula $\Theta_D = (12\pi4nR/5\beta)^{1/3}$ by using the fitted value of $\beta$, where $R$ is the gas constant and $n$ is the number of atoms per formula unit. The lattice becomes more stable when the Se-Se Van der Waals bonds were changed into Se-Cu-Se ionic bonds by increasing Cu content in $Cu_xNbSe_{2-y}S_y$ which reflected in the result Figs. 3 shows that the Debye temperature increase modestly (Table 1). Thus, we can calculate the electron-phonon coupling constant ($\lambda_{ep}$) by using the Debye temperature ($\Theta_D$) and critical temperature $T_c$ from the inverted McMillan formula:[37] $\lambda_{ep} = \frac{1.04 + \mu^* \ln\left(\frac{\Theta_D}{1.45T_C}\right)}{(1-1.62\mu^*)\ln\left(\frac{\Theta_D}{1.45T_C}\right) - 1.04}$ the values of $\lambda_{ep}$ are 0.80 for $Cu_{0.02}NbSe_{1.98}S_{0.02}$ (**Table 1**). This value suggests strong coupling superconductivity properties. The electron density of states at the Fermi level ($N(E_F)$) can be calculated from $N(E_F) = \frac{3}{\pi^2 k_B^2 (1+\lambda_{ep})}\gamma$ with the $\gamma$ and $\lambda_{ep}$. This yields value that $N(E_F)$ = 4.08 states/eV f.u. for NbSe$_2$ and $N(E_F)$ = 3.62 states/eV f.u. for Cu$_{0.02}$NbSe$_{1.98}$S$_{0.02}$ (**Table 1**). The results indicate that the density of electronic states at the Fermi energy thus decreases obviously with increasing Cu content into 2H-NbSe$_2$. With the goal of determining the critical fields $\mu_0H_{c2}(0)$, we further examined temperature dependent electrical resistivity under applied magnetic for selected Cu$_x$NbSe$_{2-y}$S$_y$ samples. **Fig. 4** reveals the $\rho(T,H)$ data measured for Cu$_x$NbSe$_{2-y}$S$_y$ ($x$ = 0.01, 0.02) which was shown as an example. The inset of **Figs. 4A and B** show upper critical field values $\mu_0H_{c2}$ plotted *vs.* temperature with $T_c$s obtained from resistively different applied fields. The solid lines through the data show the nicely linear fitting for $\mu_0H_{c2}$ *vs. T* curve near $T_c$ of two selected samples. The value of fit data slopes (d$H_{c2}$/d$T$) of selected samples are shown in **Table 1**. We can estimate the zero-temperature upper critical fields (upper inset of **Figs. 4A and B**) to 3.5 T for Cu$_{0.01}$NbSe$_{1.99}$S$_{0.01}$ and 3.0 T for Cu$_{0.02}$NbSe$_{1.98}$S$_{0.02}$ from these data, using the Werthamer-Helfand-Hohenberg (WHH) expression for the dirty limit superconductivity, $\mu_0H_{c2}$ = -0.693$T_c$ (d$H_{c2}$/d$T_c$). [37-41] Also the results obtained were summarized in **Table 1**. The Pauli limiting field for Cu$_x$NbSe$_{2-y}$S$_y$ ($x$ = 0.01, 0.02) was calculated from $\mu_0H^P$ = 1.86$T_c$. The calculated values of $\mu_0H^P$ were larger than the estimated values. Then, with this formula $\mu_0H_{c2} = \frac{\phi_0}{2\pi\xi_{GL}^2}$, where $\phi_0$ is the flux quantum, the Ginzburg-Laudau coherence length ($\xi_{GL}(0)$) was calculated ~ 9.7 nm for Cu$_{0.01}$NbSe$_{1.99}$S$_{0.01}$, and ~ 10.3 nm for Cu$_{0.02}$NbSe$_{1.98}$S$_{0.02}$ (**Table 1**).

STM imaging gives rise to the real space electronic state information, which has been proved to be the direct experimental evidence of a CDW phases. The Cu$_{0.06}$NbSe$_{1.71}$S$_{0.08}$ sample was characterized by our low-temperature STM. **Fig. 5** presents the STM image, which clearly displays the formation of the 2 × 2 commensurate CDW on the sample's surface. (The voltage dependent STM images are shown in SI **Figure S1**). We thus confirmed the formation of CDW on the sample.

Finally, we got the superconductivity phase diagram plotted $T_c$s *vs* doping content for 2H-Cu$_x$NbSe$_{2-y}$S$_y$ (0 ≤ $x$ = $y$ ≤ 0.1) was summarized in **Fig. 6**. As comparison, the information for $Cu_xNbSe_2$, $Fe_xNbSe_2$ and $NbSe_{2-x}S_x$ ($0 \leq x \leq 0.1$) were all taken from those previous literatures. The $T_c$s extracted from the resistivity measurements performed here for $Cu_xNbSe_{2-y}S_y$; the $x$ dependence of $T_c$. It's not surprising that the $T_c$ of sulfur doped material of 2H-NbSe$_2$ presents a very small change compared to the $T_c$ of pure NbS$_2$ (2H structure) of 6.5 K. In contrast, intercalation of magnetic ions may lead to a sharp drop of $T_c$, so that the $T_c$s decrease very rapidly with increasing Fe content in $Fe_xNbSe_2$. However, nonmagnetic chemically Cu intercalation was found to act somewhere between the Fe- and S-doped extremes in $T_c$ vs x curves, with a previously unknown example of an *S*-shaped suppression of $T_c$ by substitution or doping in a single-phase material. This phenomenon also exists in Cu- and S-doping of 2H-NbSe$_2$ ($Cu_xNb_{1-x}Se_{2-y}S_y$). Thus, when we adopt Cu-intercalation and S-substitution in 2H-NbSe$_2$ simultaneously, the $T_c$ decrease light rapidly and the leveling off appear at lager x content compared with $Cu_xNbSe_2$. $Cu_xNbSe_{2-y}S_y$ materials may provide a new platform for our understanding of multiband superconductivity phenomena and CDW in TMDs

## Acknowledgements

H. X. Luo acknowledges the financial support by "Hundred Talents Program" of the Sun Yat-Sen University and Natural Science Foundation of China (21701197). H. Zheng, J. Ma, Y.H. Wang are supported by the "One Thousand Youth Talents" Program of China. M.R. Li is supported by the he "One Thousand Youth Talents" Program and the National Natural Science Fund of China (NSFC-21875287). H. Zheng is supported by the National Natural Science Foundation of China (Grant No. 11674226, 1790313); National Key Research and Development Program of China (Grant No. 2016YFA0300403). J. Ma is supported by the National Natural Science Foundation of China (Grant No. 11774223). We are grateful to Weiwei Xie for fruitful discussions.

**Table 1. Characterization of the superconductivity in the $Cu_xNbSe_{2-y}S_y$ family.**

| $x = y$ in $Cu_xNbSe_{2-y}S_y$ | 0 | 0.01 | 0.02 | 0.04 | 0.05 | 0.06 | 0.07 | 0.08 | 0.09 |
|---|---|---|---|---|---|---|---|---|---|
| $T_c$ (K) | 7.16 | 6.78 | 6.12 | 4.65 | 3.93 | 3.77 | 2.91 | 2.34 | 2.26 |
| $\gamma$ (mJ mol$^{-1}$ K$^{-2}$) | 17.4(20) | - | 15.31(5) | | - | | - | - | - |
| $\beta$ (mJ mol$^{-1}$ K$^{-4}$) | 0.56 | - | 0.55 | | - | | - | | - |
| $\Theta_D$ (K) | 218(16) | - | 220(2) | | -- | | - | - | - |
| $\Delta C/T_c$ | 2.04 | - | 2.16 | | - | | - | - | - |
| $\lambda_{ep}$ | 0.81 | - | 0.80 | | - | | | - | - |
| $N(E_F)$ (states/eV f.u) | 4.08 | - | 3.62 | | - | | - | - | - |
| $-dH_{c2}/dT$ (T/K) | 1.95(4) | 0.735 (2) | 0.713 (5) | - | - | | - | - | - |
| $\mu_0H_{c2}$(T) | 9.7(2) | 3.5(3) | 3.0(2) | - | - | | - | - | - |
| $\mu_0H^P$(T) | 13.2 | 12.6 | 11.3 | 8.6 | 7.3 | 7.0 | 5.4 | 4.3 | 4.2 |
| $\xi_{GL}(0)$ (nm) | 5.8 | 9.7 | 10.3 | - | - | | - | - | - |

# Figures legends

**Figure 1. Structural and chemical characterization of $Cu_xNb_{1-x}Se_{2-y}S_y$ and $Cu_xNbSe_{2-y}S_y$.** (A) Powder XRD patterns (Cu Kα) for the $Cu_xNb_{1-x}Se_{2-y}S_y$ samples studied ($0 \leq x \leq 0.1$). Inset shows the enlargement of peak (002). (B) Powder XRD patterns (Cu Kα) for the $Cu_xNbSe_{2-y}S_y$ samples studied ($0 \leq x \leq 0.1$). Inset shows the enlargement of peak (002). (C) Powder XRD pattern with Rietveld refinement for $Cu_{0.04}Nb_{0.96}Se_{1.96}S_{0.04}$. Inset shows the crystal structure of 2H-NbSe$_2$ with Cu- and S-doping. (D) Powder XRD pattern with Rietveld refinement for $Cu_{0.04}NbSe_{1.96}S_{0.04}$. Inset shows the crystal structure of 2H-NbSe$_2$ with Cu-intercalation and S-doping. (E) The evolution of lattice parameter $a$ of $Cu_xNbSe_{2-y}S_y$ and $Cu_xNb_{1-x}Se_{2-y}S_y$. (F) The evolution of lattice parameter $c$ of $Cu_xNbSe_{2-y}S_y$ and $Cu_xNb_{1-x}Se_{2-y}S_y$.

**Figure 2. Transport characterization of the normal states and superconducting transitions for $Cu_xNbSe_{2-y}S_y$ and $Cu_xNb_{1-x}Se_{2-y}S_y$.** (A) Magnetic susceptibilities for $Cu_xNbSe_{2-y}S_y$ ($0 \leq x = y \leq 0.06$) at the superconducting transitions; applied DC fields are 20 Oe. (B) Magnetic susceptibilities for $Cu_xNbSe_{2-y}S_y$ ($0 \leq x = y \leq 0.1$) at the superconducting transitions; applied DC fields are 20 Oe. (C, D) The temperature dependence of the resistivity ratio ($\rho/\rho_{300K}$) for polycrystalline $Cu_xNbSe_{2-y}S_y$ ($0 \leq x = y \leq 0.1$). Inset of (C) shows metallic temperature dependence ($d\rho/dT$) in the temperature region of 2 - 8 K.

**Figure 3. Heat Capacity characterization of $Cu_{0.02}NbSe_{1.98}S_{0.02}$.** Debye temperature of $Cu_{0.02}NbSe_{1.98}S_{0.02}$ obtained from fits to data in applied field. Inset show heat capacities through the superconducting transitions without applied magnetic field for $Cu_{0.02}NbSe_{1.98}S_{0.02}$.

**Figure 4. Characterization of the critical fields of $Cu_xNbSe_{2-y}S_y$.** (A, B): Low temperature resistivity at various applied fields for the examples of $Cu_{0.01}NbSe_{1.99}S_{0.01}$ and $Cu_{0.02}NbSe_{1.98}S_{0.02}$

**Figure 5.** A STM image (15 × 15nm$^2$, 0.1V, 2nA) and its corresponding Fourier transformed image on a $Cu_{0.06}NbSe_{1.71}S_{0.08}$ surface. Both the patterns in real space and reciprocal space clear show the appearance of the 2 × 2 charge density waves. In the Fourier transformed image, the red (black) circle marks the one-fold (two -fold) lattice.

**Figure 6.** The superconducting phase diagram for 2H-$Cu_xNbSe_{2-y}S_y$ and 2H-$Cu_xNb_{1-x}Se_{2-y}S_y$ compared to those of 2H-$Cu_xNbSe_2$, 2H-NbSe$_{2-x}S_x$, and 2H-Fe$_x$NbSe$_2$(Refs.16, 24, 28).

**Fig. 1.**

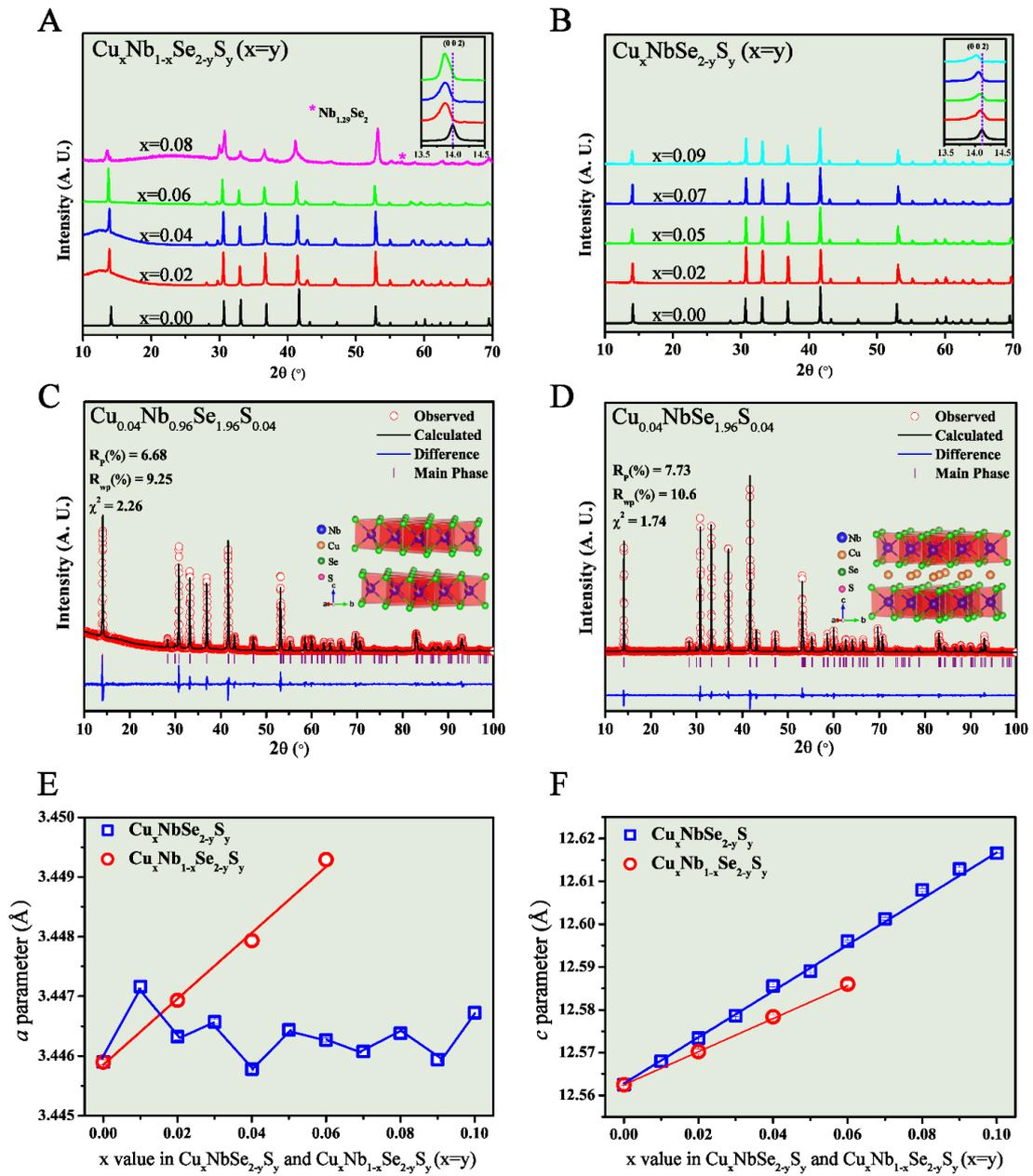

**Fig. 2.**

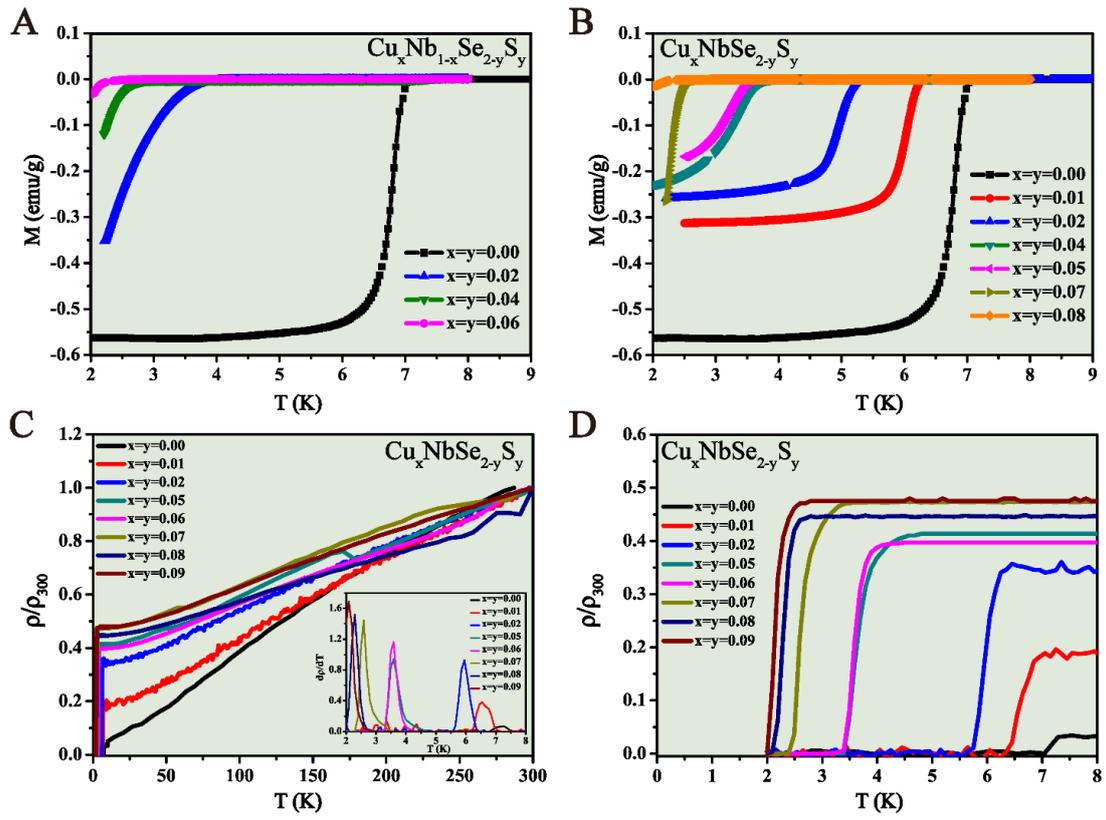

**Fig. 3.**

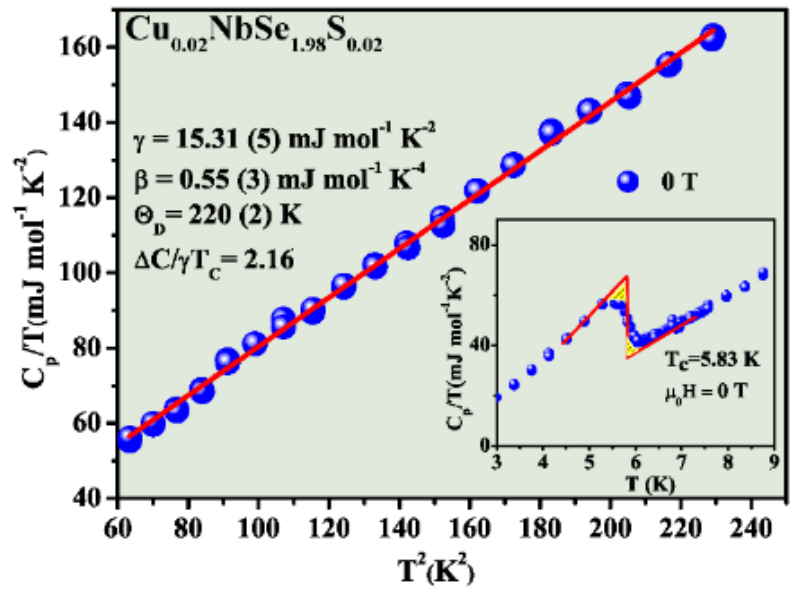

**Fig. 4.**

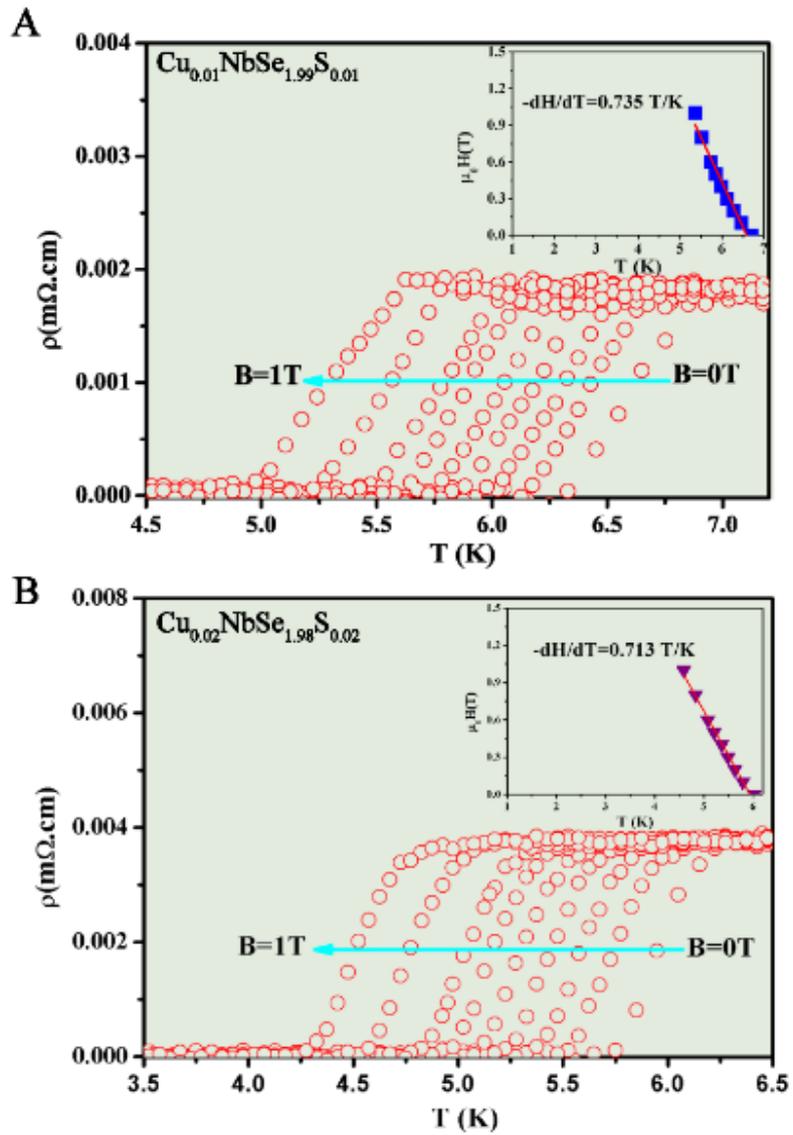

**Fig. 5.**

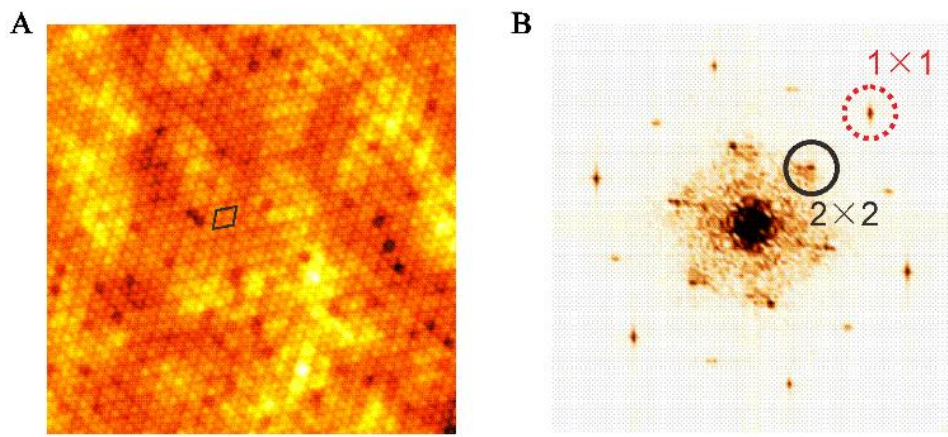

**Fig. 6.**

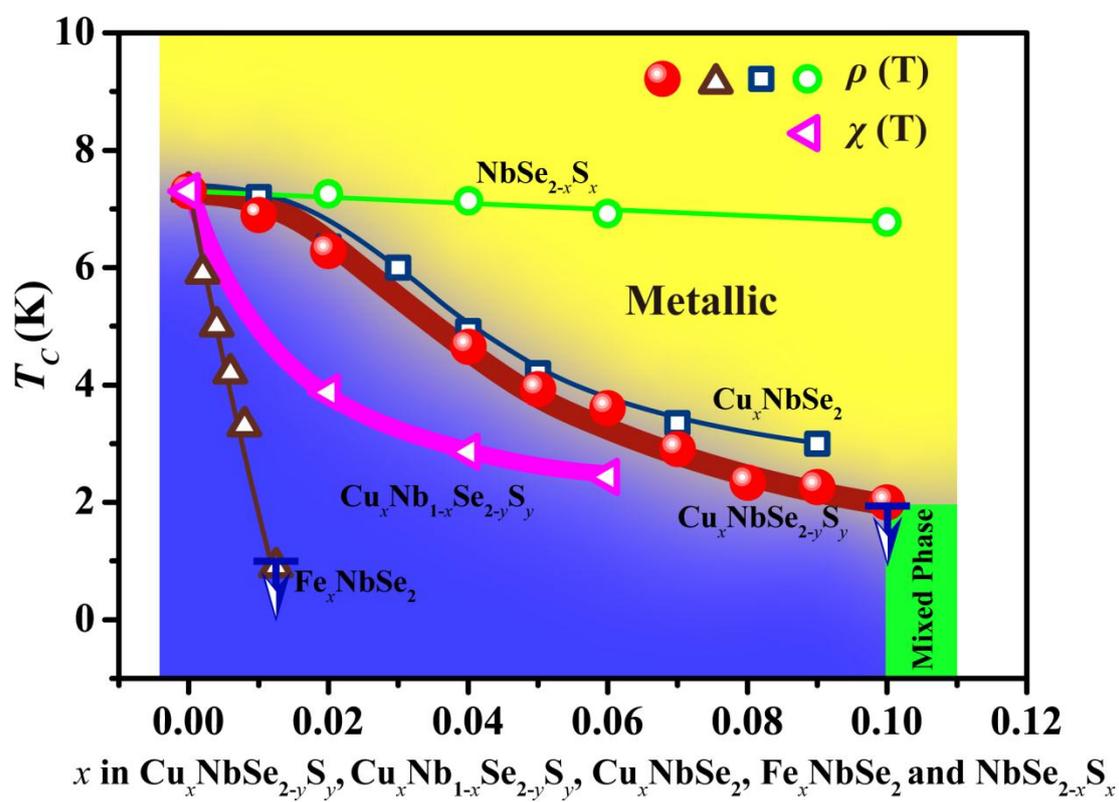

# The Unusual Suppression of Superconducting Transition Temperature in Double-Doping 2H-NbSe$_2$

*Dong Yan[1], Yishi Lin[2], Guohua Wang[3], Zhen Zhu[3], Shu Wang[1], Lei Shi[1], Yuan He[1], Man-Rong Li[4], Hao Zheng[3], Jie Ma[3], Jinfeng Jia[3], Yihua Wang[2], Huixia Luo[1]\**

[1]School of Material Science and Engineering and Key Lab Polymer Composite & Functional Materials, Sun Yat-Sen University, No. 135, Xingang Xi Road, Guangzhou, 510275, P. R. China
[2]Department of Physics, Fudan University, Shanghai, 200433, China
[3]School of Physics and Astronomy, Shanghai Jiao Tong University, Shanghai 200240, China
[4]School of Chemistry, Sun Yat-Sen University, No. 135, Xingang Xi Road, Guangzhou, 510275, China

E-mail: luohx7@mail.sysu.edu.cn

**Supporting Information**

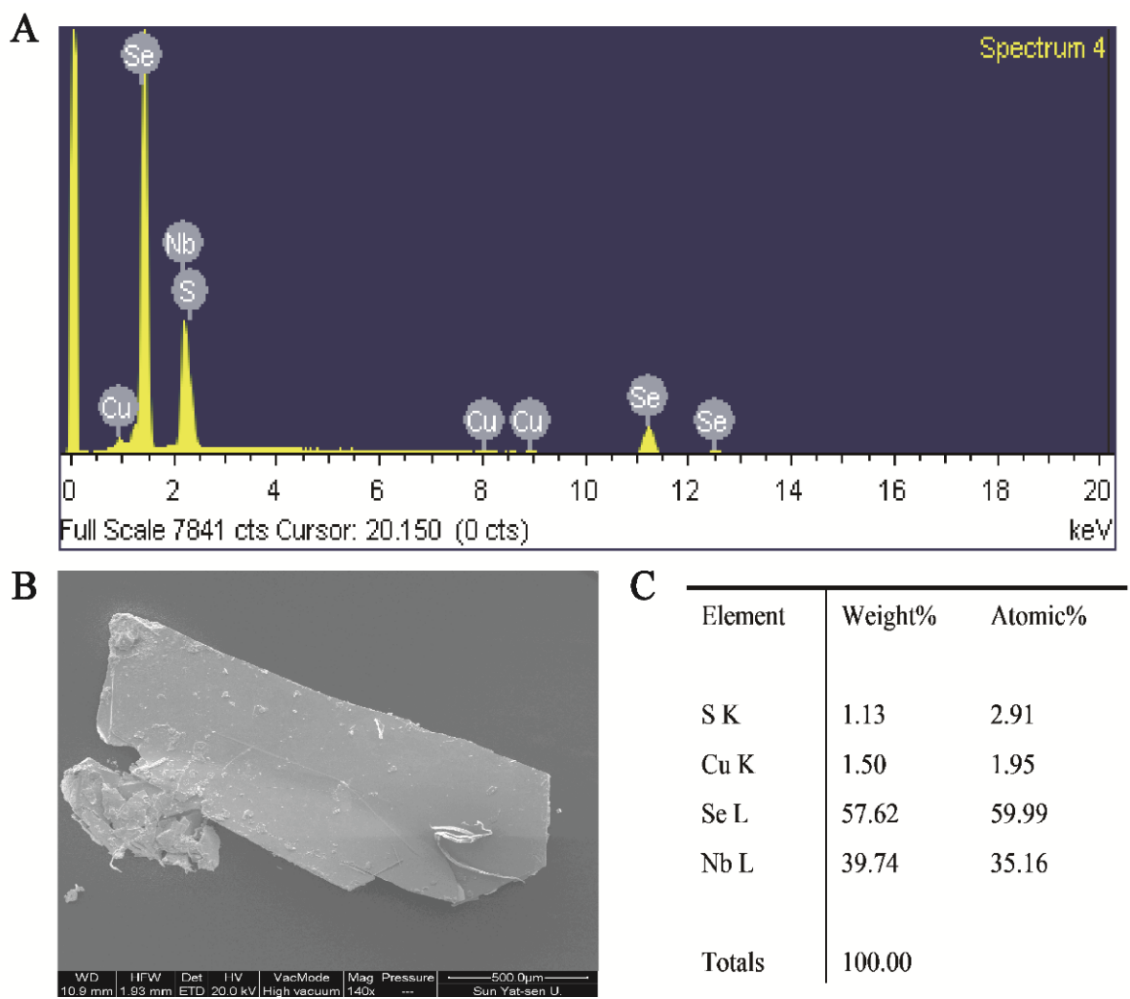

**Figure S1.** SEM and EDXS tests on the single crystals of $Cu_xNbSe_{2-y}S_y$. (A) EDXS spectrum of $Cu_{0.06}NbSe_{1.71}S_{0.08}$. (B) SEM image of $Cu_{0.06}NbSe_{1.71}S_{0.08}$ in the magnification of 140. (C) Element ratio of $Cu_{0.06}NbSe_{1.71}S_{0.08}$.

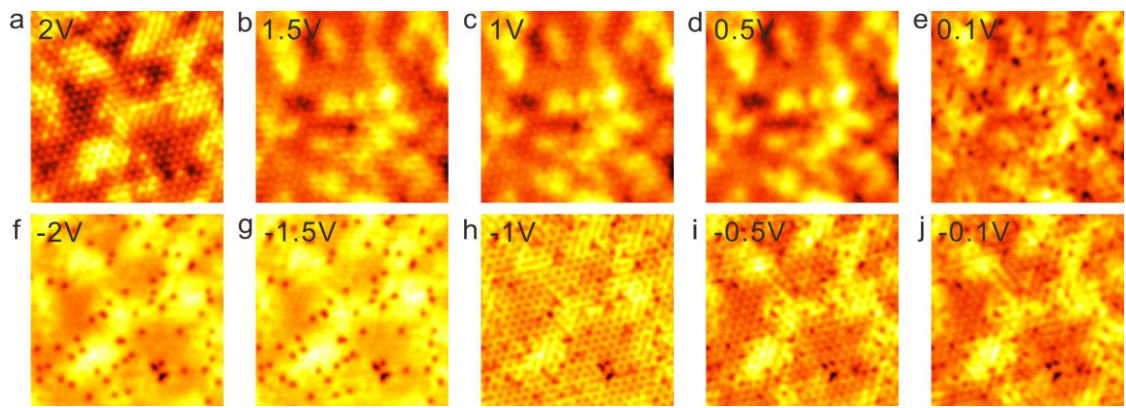

**Figure S2.** The voltage dependent STM images (all sizes of 15 x 15 nm$^2$) on a Cu$_{0.06}$NbSe$_{1.71}$S$_{0.08}$ surface. Voltages are indicated on the figures. Tunneling current is 1nA (a-c, f-j) and 0.1nA (d and e), respectively. Panel a, f-j are measured on same area, while b-e are acquired on anthers area.